\documentstyle[editedvolume,epsf]{crckapb} 
\def\lsim{\lower.5ex\hbox{$\; \buildrel < \over \sim \;$}}
\def\gsim{\lower.5ex\hbox{$\; \buildrel > \over \sim \;$}}
\input psfig.tex

\newcommand{\be}{\begin{equation}}
\newcommand{\ee}{\end{equation}}
\def\lsim{\lower.5ex\hbox{$\; \buildrel < \over \sim \;$}}
\def\gsim{\lower.5ex\hbox{$\; \buildrel > \over \sim \;$}}

\runningtitle{Fundamental Modes of Black Hole Accretion}
\begin{opening}
\title{Fundamental States of Accretion/Jet Configuration and the Black Hole
Candidate GRS1915+105}

\author{Sandip K. Chakrabarti$^1$ and Anuj Nandi}
\institute{S. N. Bose National Centre For Basic Sciences\\
	JD Block, Salt Lake, Sector-III, Calcutta-700098, India\\
	email: chakraba@boson.bose.res.in, anuj@boson.bose.res.in}
\end{opening}

\begin{document}

\begin{abstract}
{\small Advective disk paradigm of black hole accretion
includes self-consistent formation of shocks and outflows from 
post-shock region. We apply this paradigm to understand rich variation
of the light curve of the black hole candidate GRS1915+105.
We propose that out of five possible {\it fundamental states}
the black hole candidate GRS1915+105 moves around among three of them
creating all possible observed light curves.}
\end{abstract}
\smallskip



\noindent $1$ Also, Honorary Scientist at Centre for Space Physics, 
IA-212, Salt Lake City, Sector-III, Calcutta-700097\\

\noindent Submitted October, 2000 to Indian Journal of Physics; In press.

\section{Introduction}

Chakrabarti [1-2] pointed out that considerable amount of outflow
could be generated from the centrifugal pressure supported boundary layer (CENBOL) of a
black hole. Indeed, it was shown that when the shock is weak (compression ratio $R\sim 1$), 
the outflow must be negligible and when the shock is strong ($R\sim 4-7$),
the outflow is small but non-negligible. However, for the intermediate
shock strength ($R\sim 2-3$) the outflow rate is very large -- close to 
thirty percent of the inflow rate. Subsequently, Chakrabarti [3]
showed that the slope of the hard-tail of the spectrum of a black hole must 
become larger in presence of outflows from the CENBOL region and conversely,
if external matter is added to CENBOL for the same intensity of soft photons
(from the Keplerian disk), the spectral slope must  become smaller.  In other words,
hard-state spectrum should be softened and soft-state spectrum should be hardened. This 
has also been observed to be the case [4].

Another important phenomenon involving outflow is its periodic cooling by Compton scattering.
When the outflow rate is large, the slowly moving subsonic region could be catastrophically
cooled down by soft photons from the Keplerian disk [3,5]. 
The sonic surface of the cooler outflow comes closer to the black hole horizon and
the flow separates into two parts. Matter from the region above the new sonic sphere
separates supersonically as blobs, and matter below the new sonic sphere
returns back to the accretion disk. This causes enhancement of 
accretion rates of the disk temporarily in a very short time-scale and could produce
interesting temporal variation of the photon flux. 

Meanwhile, Belloni et al. [6] and Nandi et al. [7]
classified all possible types of light curves 
of a very exciting black hole candidate GRS1915+105. B2000 divided 
the light curves in twelve types (termed as $\phi$, $\kappa$, $\gamma$, $\mu$, $\delta$,
$\theta$, $\lambda$, $\kappa$, $\rho$, $\nu$, $\alpha$ and $\beta$) 
and N2000 divided the light curves in four fundamental classes (Hard, Soft, 
Semi-Soft and Intermediate). Belloni et al. [6] mentioned that from the spectral point 
of view, however, one could imagine that there are three types of States: 
A, B and C combining which these light curves could be generated. 
In the present {\it Rapid Communication}, we claim that existence of these states
{\it fundamental states} cannot be understood by a standard Keplerian 
disk model and can be easily understood from the advective disk paradigm. 
In the next Section, we present the backbone solutions of the advective disk.
In \S 3, we discuss five fundamental states which are formed out of these
backbone solution and show that the states A, B and C of the black hole
candidate GRS1915+105 [6] comprise three of them. 
Finally in \S 4, we draw our conclusions.

\section{Backbone solutions of Advective Accretion}

Equations governing advective accretion disks in pseudo-Newtonian geometry and in Kerr geometry 
were presented elsewhere and will not be repeated here [8-9].
It is observed that there are a total of eight different types of solutions [9]
which are denoted as: \\

\noindent O: Flow passing through the outer sonic point only.\\
\noindent I: Flow passing through the inner sonic point only.\\
\noindent SA: Flow has two saddle type sonic points, and  a steady shock forms in accretion.\\
\noindent NSA: Same as SA but no steady shocks can form. Shocks are oscillatory.\\
\noindent SW: Same as SA but steady shocks form only in winds.\\
\noindent NSW: Same as SW but oscillatory shocks form in winds.\\
\noindent $I^*$: Incomplete solution with inner sonic point.\\
\noindent $O^*$: Incomplete solution with outer sonic point.\\

In Fig. 1, one solution (in Kerr geometry) from each type is shown in Mach 
number (y-axis) vs. $log(r)$ (x-axis) plane. Vertical equilibrium 
and axi-symmetry have been assumed. For each of the solutions we also present
the schematic diagram of the nature of the flow. Transverse thickness is estimated
from the assumption of vertical equilibrium $h\sim a r^{1/2}(r-1)$, where $r$
is the radial distance in units of Schwarzschild radius $r_g=2GM_B/c^2$ ($M_B$
is the black hole mass and $G$ and $c$ are the gravitational constant and velocity 
of light respectively.). One notes that close to the black hole, matter is puffed 
up since its temperature is higher. Non-steady solutions have been 
represented by turbulence [in (g) and (h)].

When viscosity is added, closed topology of the
solutions shown above open up [8] and the flow
can join with a Keplerian disk. The specific energy of a Keplerian flow:
$$
{\cal E}= \frac{1}{2} v^2 + n {a^2} +\frac{\lambda^2}{2 r^2} -\frac{1}{2(r-1)}
\eqno{(1)}
$$
where, $n$ is the polytropic constant, $v$ is the radial velocity, $a$ is the sound 
velocity and $\lambda$ is the specific angular momentum respectively. For a cooler 
flow, ${\cal E}$ is negative, but for a hotter flow $a$, particularly away from the
equatorial plane (so that the last term in Eq. 1 is small), the energy could be 
positive. In case matter brings in magnetic field its dissipation would raise
the energy to a positive value so that shocks may form in a steady flow.
In a non-steady flow such restrictions do not apply and oscillating shocks
may form even with bound flows  [10-11].

Fig. 2 shows the representative solutions in a viscous flow and how a 
realistic disk looks like. In (a), flow viscosity is smaller than the critical value [8]
and shocks can form and outflows are produced from CENBOL as in Fig. 1. In (b),
flow viscosity is higher than the critical value [8] and there are two solutions: 
one is mostly Keplerian (optically thick) till it passes through the inner sonic
point, and the other is Keplerian farther out and passes through the outer sonic 
point. This is the optically thin branch of the solution. 

\section{Fundamental States of a Realistic Accretion Flow}

One could combine solutions in 1(a-h) and 2(a-b) to obtain realistic accretion-wind systems
around a black hole. Viscosity is considered to be high in the equatorial plane 
and smaller away from the plane. From Fig. 2(b) equatorial flow will be Keplerian closer
to the black hole, and solutions of Fig. 2(a) would cover above and below. 
There are several possibilities for reasonable parameters of a black hole accretion. We name
these states according to the way they are commonly perceived in the literature.\\

\noindent  1. {\bf Hard State:}
The accretion rate in the Keplerian component is low ${\dot M}_K \sim 0.001-0.1$
and that of the sub-Keplerian component is high ${\dot M}_s \sim 1$. The combined
sub-Keplerian flow enters into the black hole without forming a shock. Fig. 3(a) shows the
schematic diagram of the accretion-wind system. {\it Spectral signature}: Hard state without
quasi-periodic oscillation of X-rays.\\

\noindent 2. {\bf Off State:}
The accretion rates are similar as above, but viscosity is lower than the critical
$\alpha \lsim 0.01$ ($\alpha$ is the Shakura-Sunyaev [12] viscosity parameter) 
so that shocks may form. If cooling rate in the post-shock region roughly agrees with the
inflow rate, quasi-periodic oscillation of X-rays could be seen. Outflow is
produced which intercepts soft photons from Keplerian disk. Fig. 3(b) 
shows the schematic diagram. {\it Spectral signature}: hard state with 
or without QPO. With time, the spectrum can get
softer if the sonic sphere (region till the sonic point in the outflow) 
gets filled up gradually.\\

\noindent 3. {\bf Dip State:}
Keplerian accretion rates are higher ${\dot M}_K \sim 0.1-0.3$ and viscosity 
is also higher $\alpha \gsim 0.01$ or more so that shocks are weaker. 
Post-shock flow is partly cooled due to Comptonization. Outflow till the sonic
sphere has sufficient optical depth that it is cooled by Comptonization. The sonic
point comes down as sound speed goes down in this region. Flow which remains sub-sonic
with respect to this sonic sphere loses outward drive and returns back to the disk, while
the supersonic flow separates as blobs in the jets. Fig. 3(c) shows the schematic
diagram. {\it Spectral signature}: tendency towards softer state and large spectral slope.
QPO may or may not be visible as the shock is cooler (with longer cooling time scale) while 
infall time is shorter since the Keplerian disk moves inward due to larger viscosity.\\

\noindent 4. {\bf On State:}
Original flow may remain similar to above, but the return flow
enhances both Keplerian and sub-Keplerian disk rates last few hundred Schwarzschild radii.
Fig. 3(d) shows the schematic diagram. Duration of this state is the duration of drainage of 
the excess accretion from return flow. {\it Spectral signature}: softer state with high photon
flux. QPO is absent.\\

\noindent 5. {\bf Soft State:}
Accretion rate of the Keplerian component is high ${\dot M}_K\gsim 0.3 
{\dot M}_{Edd}$ and the viscosity is high enough so that Keplerian disk 
moves in all the way to the inner edge of the disk (Fig. 2b). Matter moves almost 
radially and transfers its momentum to soft photons (bulk motion
Comptonization [13]). Fig. 3(e) shows the schematic diagram. {\it Spectral signature}:
Soft state spectrum without QPO with a weak power-law hard-tail.\\

Color-Color diagrams [HR1 vs. HR2 diagrams  where HR1=b/a and HR2=c/a 
(a:2-5keV, b:5-13keV, c:13-60keV)] showed very intricate structures 
(shapes of atoll, banana, etc.) [6].  From these, Belloni et al. [6] conclude 
that there are three distinct States of GRS 1915+105:
A (low rate and low HR1, HR2), B (high rate, high HR1) and
C (low rate, low HR1, variable HR2 depending on length of the event). 
According to classifications of Nandi et al. [7], this would correspond to 
different regions in the softness ratio diagram. It seems that the State C 
exhibits QPO. State A and state B do not exhibit QPO. More interestingly, 
except for $C\rightarrow B$ transition, all other transitions of states are 
allowed. Nandi et al. [14] found evidence of QPO in some of the State A light curve. 

A comparison of the description of the States given above and the description of the
fundamental States (1-5), it seems that,
for GRS1915+105, data obtained so far suggests that that the fundamental 
states 1 and 5 are missing. States 2-4 can be identified with 
States C, A and B of Belloni et al. [6] respectively. 
One can understand a typical evolution of States in the following way:
Suppose we start with the State 2 described above.
If the accretion rate is generally increased, shock is weakened (compression ratio goes down).
There could be two types of high count states (State 3 [dip] and State 4 [On]).
After the winds of State 2 fills in the sonic sphere and cools it 
down by Comptonization, CENBOL and the region till the sonic sphere collapse. 
This is the State 3. Now there are two possibilities [5]: either the flow
separates completely as a blob and returns to State 2 or the flow mostly
returns back to the accretion disk and enhances the accretion. 
This would be the State 4. This may in turn increases the outflow [1-2].
But shock becomes weaker because of post-shock cooling, hence the outflow 
is very mild, but may remain at the threshold so that a bit more outflow can cause
the sonic sphere to collapse again. Thus, occasional trips to State 3 from State 4 is possible.
This is regularly observed [6].
Once the enhanced matter is drained out and the shock bounces back to roughly 
the original location (compatible with its specific energy and angular momentum)
State 2 forms again. Since State 2 produce fewer soft photons, State 4 is
not directly possible from State 2 without first producing return flow and
enhanced accretion. This may explain why a direct transition from 
State 2 to State 4 has not been seen so far [6].

It is clear that the complex behaviour of GRS1915+105 necessarily requires
both Keplerian and sub-Keplerian disks for a proper explanation of the 
light curve. The return-flow from the cooler wind acts as a non-linear feedback which can be
represented schematically in Fig. 4. Here, ${\dot M}_K$, ${\dot M}_S$, ${\dot M}_{in}$,
${\dot M}_{rf}$, ${\dot M}_{out}$ and ${\dot M}_{BH}$ represents Keplerian accretion rate,
sub-Keplerian accretion rate, total accretion rate, outflow rate and the
rate of actual accretion to the black hole respectively. The soft photon intensity  $S_\gamma$
intercepted by the sonic sphere and the CENBOL is a function of the Keplerian rate. 

\section{Concluding Remarks}

We have presented the fundamental states of a viscous advective disk which includes
radiative transfer. We identify that trips through States 2-4 cause the variable light curves
in GRS 1915+105. Outflow seems to be a determining factor in switching 
these states. Numerical simulations showed existence of such outflows [15]. We 
already noted that the spectral states are related to the outflow rates [5].
Observations also suggest such a possibility [16-17]. We believe that 
non-linear feed-back from the outflowing wind is
essential to understand the variable light curves observed in this black hole candidate.
Detailed modeling of these light curves would be presented elsewhere.

\section{Acknowledgments}

This project is partly supported by a DST grant No. SP/S2/K-14/98.  

{}

\newpage
\centerline{Figure Captions}

\noindent Fig. 1: All possible representative solutions of an inviscid advective flow (A-H)
and the neture of the disk-jet system (a-h). In the solutions,  Mach numbers (y-axis)
are plotted against logarithmic radial distance (x-axis).

\noindent Fig. 2: Same as Fig. 1 but for a viscous flow. (a) Viscosity smaller than the
critical value and (b) larger than the critical value.

\noindent Fig. 3: Fundamental States of the black hole accretion-jet system obtained 
by combining backbone solutions of Fig. 1 and Fig. 2. (a) Hard State, (b) Off State,
(c) Dip State, (d) On State and (e) Soft State. 

\noindent Fig. 4: Schematic diagram stressing non-linearity induced by the return flow 
from the jet to the disk. Varieties of the light curve of GRS1915+105 is expected to be generated
because of this no-linearity.

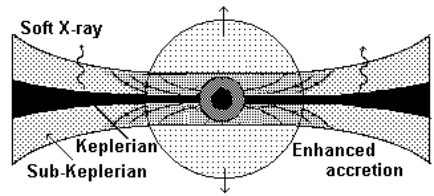
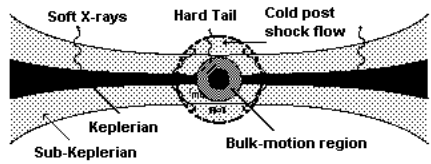
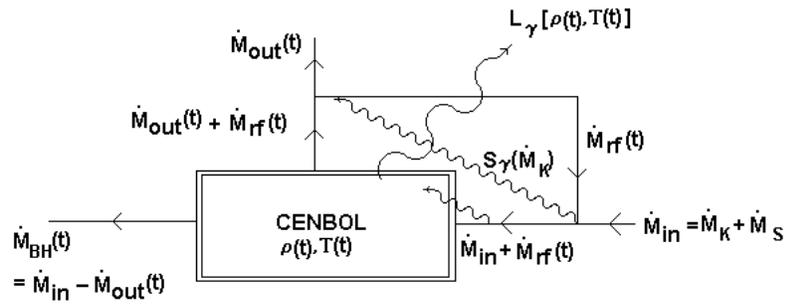
\end{document}